\documentclass[conference,10pt]{IEEEtran}
\usepackage{epsfig,rotating,setspace,latexsym,amsmath,epsf,amssymb,amsfonts,bm,theorem,subfigure,epstopdf}

\usepackage{amsbsy}
\usepackage{cite,authblk}
\usepackage{bbm}
\usepackage{color, xcolor}
\usepackage{mathtools}
\usepackage{algorithm}
\usepackage{textcomp}
\usepackage[noend]{algpseudocode}
\usepackage{verbatim}
\usepackage{graphicx, graphics}
\usepackage{afterpage}
\usepackage{placeins}
\usepackage{bm}
\usepackage{authblk}
\usepackage[symbol]{footmisc}
\usepackage{multirow}
\usepackage{booktabs}

\DeclareMathOperator*{\argmax}{arg\,max}
\DeclareMathOperator*{\argmin}{arg\,min}

\algrenewcommand\algorithmicforall{\textbf{foreach}}
\algrenewcommand\algorithmicindent{.8em}

\IEEEoverridecommandlockouts
\allowdisplaybreaks

\begin{document}

\title{Deep Learning Based Joint Multi-User MISO Power Allocation and Beamforming Design}

\author{
Cemil Vahapoglu$^{\dag}$, Timothy J. O’Shea$^{*}$, Tamoghna Roy$^{*}$, Sennur Ulukus$^{\dag}$ \\
\normalsize $^{\dag}$University of Maryland, College Park, MD, $^{*}$DeepSig Inc., Arlington, VA \\
\normalsize \emph{cemilnv@umd.edu, tim@deepsig.io, tamoghna.roy@deepsig.io, ulukus@umd.edu}
}

\maketitle

\begin{abstract}
    The evolution of fifth generation (5G) wireless communication networks has led to an increased need for wireless resource management solutions that provide higher data rates, wide coverage, low latency, and power efficiency. Yet, many of existing traditional approaches remain non-practical due to computational limitations, and unrealistic presumptions of static network conditions and algorithm initialization dependencies. This creates an important gap between theoretical analysis and real-time processing of algorithms. To bridge this gap, deep learning based techniques offer promising solutions with their representational capabilities for universal function approximation. We propose a novel unsupervised deep learning based joint power allocation and beamforming design for multi-user multiple-input single-output (MU-MISO) system. The objective is to enhance the spectral efficiency by maximizing the sum-rate with the proposed joint design framework, NNBF-P while also offering computationally efficient solution in contrast to conventional approaches. We conduct experiments for diverse settings to compare the performance of NNBF-P with zero-forcing beamforming (ZFBF), minimum mean square error (MMSE) beamforming, and NNBF, which is also our deep learning based beamforming design without joint power allocation scheme. Experiment results demonstrate the superiority of NNBF-P compared to ZFBF, and MMSE while NNBF can have lower performances than MMSE and ZFBF in some experiment settings. It can also demonstrate the effectiveness of joint design framework with respect to NNBF. 
\end{abstract}

\section {Introduction}
Wireless physical layer research has broadly focused on waveform design, signal detection and estimation techniques, and channel characterization. This includes tasks such as interference management, transceiver chain design, error-correcting algorithms design to provide reliable data transfer \cite{OsheaDL}. With the advancement of fifth-generation (5G) wireless networks, there is an increased demand for high data rate, high spectral efficiency, extensive coverage, low latency, and power efficiency. These issues of concern can be evaluated within the scope of wireless resource management problems, which span various domains such as spectrum management, cache management, computation resource management, power control, transmit/receive beamforming design \cite{AppML}.

Traditional wireless communication system designs and implementations require strong probabilistic modeling and signal processing techniques \cite{erpek2020deep}. However, they have challenging limitations in terms of computational complexity due to rigorous computations, which creates an important gap between theoretical analysis and real-time processing of algorithms. In addition to the substantial computational complications, many of the existing designs are non-practical for dynamic network scenarios by producing suboptimal results with their presumptions of static network conditions, and dependencies in algorithm initialization \cite{2011iterativeWeightedMMSE,2008_WSR_WMMSE}.

On the other hand, machine learning (ML) presents robust automated systems capable of learning from dynamic spectrum data, rather than relying on solely policy based solutions for specific scenarios \cite{ClancyCognitive}. Recent advancements in powerful graphical processing units (GPUs) and the exponential growth of available data and compute have particularly empowered deep learning based methods, enabling them to attain considerable representational capabilities \cite{AppML}. Furthermore, it has been proven that the deep neural networks (DNN) offer universal function approximation for conventional high complexity algorithms. Therefore, DNNs can also be utilized for numerical optimization problems addressing wireless resource management problems such as beamforming design and power control, which can be treated as nonlinear mapping functions to be learned by the DNNs \cite{Sun2018}.

In this paper, we focus on transmit (often downlink) beamforming design and transmit beamforming power control, which are significant challenges in 5G wireless communication networks. In the literature, numerous deep learning based beamforming design methods have been proposed for different multiple antenna configurations. \cite{Zhang_2023} proposes a joint learning framework for channel prediction, transmit beamforming prediction, and power optimization in multi-user multiple-input single-output (MU-MISO) setting. However, it utilizes the parameterized structure of beamforming solution given the power values for sum-rate maximization suggested by \cite{OptMultiUserTransmitBeamforming}, rather than offering an end-to-end beamforming design. \cite{DeepTx2022} proposes a convolutional neural network (CNN) architecture for downlink transmit beamforming design by utilizing uplink channel estimate in a supervised manner. Additionally, \cite{WenchaoMISODownlinkBF} proposes deep learning frameworks for signal-to-interference-plus-noise ratio (SINR) balancing problem, power minimization problem, and sum-rate maximization problem. For sum-rate maximization, they also utilize the optimal beamforming structure suggested by \cite{OptMultiUserTransmitBeamforming}. It involves matrix inversion operations, which can create computational burden for a real-time processing system for massive multiple-input multiple-output (mMIMO) systems. Furthermore, semi-supervised learning is employed for power allocation to maximize sum-rate in \cite{WenchaoMISODownlinkBF}, which can be non-practical in the case of unavailability of annotated data.

In our work, we propose a novel deep learning based joint power allocation and beamforming design for MU-MISO setting. The proposed framework is denoted as NNBF-P when NNBF represents end-to-end beamforming design assuming equal transmit power for all user equipment (UE) without power allocation. The proposed framework performs unsupervised training. To the best of our knowledge, it is the first work that utilizes unsupervised DL training for a joint power allocation and beamforming design scheme, targeting the sum-rate maximization problem. We conduct the performance analysis of proposed framework by comparing with zero-forcing beamforming (ZFBF) technique and minimum mean square error (MMSE) beamforming, which are considered as our baselines. Additionally, we compare it with NNBF to evaluate the advantage of power allocation. Spectral efficiency is considered as performance metric while the computational efficiency compared to ZFBF and MMSE has been shown previously \cite{vahapoglu2023deep}. Experimental results demonstrate the superiority of the proposed framework compared to ZFBF, MMSE, and NNBF. Furthermore, NNBF remains inferior relative to MMSE and comparable with ZFBF for some experiment settings. Experimental results also show the success of the joint design framework with respect to NNBF.

\section{System Model \& Problem Formulation}
\subsection{Downlink Multi-User MISO (MU-MISO) Setup}

\begin{figure}[h]
 \centerline{\includegraphics[width= 1\linewidth]{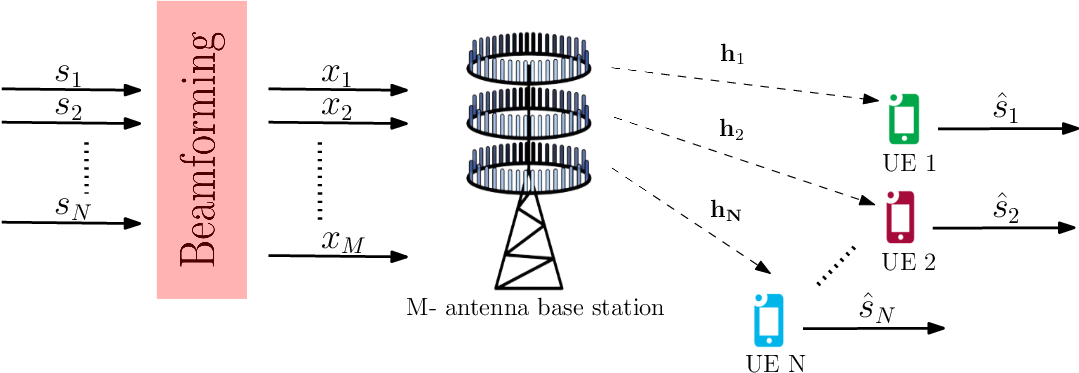}}
  \caption{Downlink massive MIMO where BS transmit data streams to single-antenna UEs with allocated powers on the same time/frequency resources.}
  \label{Downlink System Model}
  \vspace*{-0.1cm}
\end{figure}

We consider a downlink transmission scenario where a base station (BS) is equipped with $M$ transmit antennas to convey $N$ data streams to $N$ single-antenna UEs as shown in Fig.~\ref{Downlink System Model}.

The downlink channel matrix is denoted as $\mathbf{H} = 
[\mathbf{h}_1  ~ \mathbf{h}_2 ~ \cdots ~ \mathbf{h}_N] \in \mathbb{C}^{M \times N}$, where $\mathbf{h}_k$ corresponds to the channel vector between UE $k$ and the BS. Downlink channel estimate is obtained by assuming the channel reciprocity between uplink channel and downlink channel. Then, we assume that downlink channel state information (CSI) is available at the BS.

Let $s_i \in \mathbb{C}$ represent the data stream to be transmitted to UE $k$, $i=1, \ldots, N$. The transmitted signal $\mathbf{x} \in \mathbb{C}^M$ can be written as
\begin{align}
    \mathbf{x} = \mathbf{W}\mathbf{s} = \sum_{i=1}^{N} \mathbf{w}_i s_i
\end{align}
where $\mathbf{s} = [s_1  ~ s_2 ~ \cdots ~ s_N]^T \in \mathbb{C}^{N}$ and $\mathbf{s}^H\mathbf{s}=1$. The downlink beamforming matrix $\mathbf{W}$ can be represented as
\begin{align}\label{dl-bf-notation}
\mathbf{W} &=[\mathbf{w}_1  ~ \mathbf{w}_2 ~ \cdots ~ \mathbf{w}_N] \nonumber\\
&=[\sqrt{p}_1\tilde{\mathbf{w}}_1  ~ \sqrt{p}_2\tilde{\mathbf{w}}_2 ~ \cdots ~ \sqrt{p}_N\tilde{\mathbf{w}}_N] \nonumber \\
&= \sqrt{\mathbf{p}} \odot \Tilde{\mathbf{W}} \in \mathbb{C}^{M \times N}
\end{align}
where $\mathbf{p} =[p_1 ~ p_2 ~ \cdots p_N]^T  \in \mathbb{R}^N$ and $\odot$ represents the elementwise multiplication. For UE $k$, $\mathbf{w}_k = \sqrt{p}_k \tilde{\mathbf{w}}_k \in \mathbb{C}^{M}$ represents the linear beamforming filter with transmit power $p_k = \mathbf{w}_k^H \mathbf{w}_k$ where $\tilde{\mathbf{w}}_k$ is the normalized beamforming filter, i.e., $\tilde{\mathbf{w}}_k^H \tilde{\mathbf{w}}_k=1$, $k=1,\ldots,N$. The total power constraint is considered as $\mathrm{tr}(\mathbf{W}^H\mathbf{W})= \sum_{i=1}^N p_i = N$.

The received signal $y_k \in \mathbb{C}$ for UE $k$, $\forall k=1,\ldots,N$ can be written as 
\begin{align}\label{received_signal}
    y_k &= \mathbf{h}_k^T\mathbf{x} + n_k = \mathbf{h}_k^T \mathbf{w}_k s_k + \sum_{i=1, i\neq k}^N \mathbf{h}_k^T\mathbf{w}_i s_i + n_k \nonumber \\
    &= \underbrace{ \sqrt{p_k} \mathbf{h}_k^T \tilde{\mathbf{w}}_k s_k }_{desired \, signal} + \underbrace{ \sum_{i=1, i\neq k}^N \sqrt{p}_i \mathbf{h}_k^T \tilde{\mathbf{w}}_i s_i}_{interfering \, signal} +\underbrace{n_k}_{noise}
\end{align}
where $\mathbf{n} = [n_1  ~ n_2 ~ \cdots ~ n_N]^T \in \mathbb{C}^{N}$ denotes the additive white Gaussian noise (AWGN) with i.i.d.~entries $n_k \sim \mathcal{CN}(0, \sigma^2), \, k=1,\ldots, N$.

\subsection{Joint Power Allocation and Beamforming Design for Sum-Rate Maximization}
Our objective is the joint design of downlink transmit power allocations and beamforming weights to maximize the sum-rate of all UEs under total power constraint $P_\textrm{max}$. $P_\textrm{max}$ is considered as $N$ throughout this work. Using the received signal for UE $k$ in (\ref{received_signal}), SINR for UE $k$ is written as
\begin{align}\label{sinr}
    \gamma_k = \frac{p_k|\mathbf{h}_k^T \mathbf{\tilde{w}}_k|^2}{\sum_{i=1, i\neq k}^N p_i|\mathbf{h}_k^T\mathbf{\tilde{w}}_i|^2 + \sigma^2}
\end{align}
Therefore, the optimization problem of interest is
\begin{align}\label{sum-rate maximization problem}
    \mathbf{\Tilde{W}}^* , \mathbf{p}^* = \argmax_{\Tilde{\mathbf{W}},\mathbf{p}} & \quad \sum_{i=1}^N \alpha_i \log(1 + \gamma_i) \nonumber \\
    \textrm{s.t.} &\quad \textrm{tr}(\mathbf{W}^H\mathbf{W}) = \sum_i^N p_i \leq P_\textrm{max}
\end{align}
where $\alpha_i$ denotes the rate weight for UE $i$.

\subsection{Optimal Multi-User Transmit Beamforming Structure for Sum-Rate Maximization}

It should be noted that the problem formulation in (\ref{sum-rate maximization problem}) is non-convex. In \cite{OptMultiUserTransmitBeamforming}, it is stated that the solution to the power minimization problem under SINR constraint in (\ref{power_minimization_problem}) must satisfy the power constraints of sum-rate maximization problem in (\ref{sum-rate maximization problem}) when it is supposed that SINR constraints are set to the optimal values of (\ref{sum-rate maximization problem}) as $\{\gamma_1^*, \ldots, \gamma_N^*\}$ since it finds the beamforming vectors given SINR values with minimal power values. Further details can be seen in \cite{OptMultiUserTransmitBeamforming}. We are only interested in the optimal multi-user transmit beamforming structure for sum-rate maximization,
\begin{align}\label{power_minimization_problem}
   \mathbf{W}^* = \argmin_{\mathbf{W}} & \quad  \sum_i^N \| w_i\|^2 \nonumber \\
   \textrm{s.t.} &\quad \gamma_i \geq \rho_i \quad \forall i=1,\ldots,N.
\end{align}

As result of connection between problems (\ref{sum-rate maximization problem}) and (\ref{power_minimization_problem}), the optimal beamforming structure is shown as \cite{OptMultiUserTransmitBeamforming} 
\begin{align} \label{optimal_structure}
    \mathbf{W}^* &= [\mathbf{w}^*_1 \mathbf{w}^*_2 \cdots \mathbf{w}^*_N] \nonumber \\
    &=[\mathbf{\tilde{w}^*}_1 \mathbf{\tilde{w}^*}_2 \cdots \mathbf{\tilde{w}^*}_N] \begin{bmatrix}\sqrt{p}^*_1 & & \\ & \ddots & \\ & & \sqrt{p}^*_N\end{bmatrix} \nonumber \\
    &= \left( \mathbf{I}_M + \frac{1}{\sigma^2} \mathbf{H}\mathbf{\Lambda^*}\mathbf{H}^H\right)^{-1} \mathbf{H}\mathbf{P^*}^{\frac{1}{2}}
\end{align}
where$\mathbf{P}^*$ is diagonal scaled optimal downlink transmit powers.

\begin{align}
    \mathbf{P}^* &= \begin{bmatrix}\frac{p_1^*}{\left( \mathbf{I}_M + \frac{1}{\sigma^2} \mathbf{H}\mathbf{\Lambda^*}\mathbf{H}^H\right)^{-1} \mathbf{h}_1} & & \\ & \ddots & \\ & & \frac{p_N^*}{\left( \mathbf{I}_M + \frac{1}{\sigma^2} \mathbf{H}\mathbf{\Lambda^*}\mathbf{H}^H\right)^{-1} \mathbf{h}_N}\end{bmatrix}
\end{align}
and $\mathbf{\Lambda^*}$ is the diagonal optimal Lagrange multipliers referred as virtual optimal uplink power allocations. It can be computed by fixed point equations \cite{OptMultiUserTransmitBeamforming}
\begin{align}
    \mathbf{\Lambda^*} &= \begin{bmatrix}\lambda^*_1 & & \\ & \ddots & \\ 
    & &\lambda^*_N\end{bmatrix}
\end{align}
    




However, finding optimal $\{p_i^*\}_{i=1}^N$ and $\{\lambda^*_i \}_{i=1}^N$ to maximize sum-rate is a non-convex problem. Locally optimal solutions can be obtained via iterative algorithms \cite{2011iterativeWeightedMMSE, 2008_WSR_WMMSE}. Locally optimal solutions can be non-practical, e.g., allocating total power to only one UE with best channel quality to maximize sum-rate, depending on initialization. In spite of suboptimality of power allocations, optimal beamforming structure $\{\mathbf{\tilde{w}}_i \}_{i=1}^N$ can be computed by (\ref{optimal_structure}). Yet, it can create large computation burden due to matrix inversion operations for massive MIMO systems. In this respect, we propose a DL framework to have end-to-end beamforming design $\{\tilde{w_i} \}_{i=1}^N$ and power allocations $\{p_i\}_{i=1}^N$ for sum-rate maximization without need of a large computational burden.

\section{Proposed Deep Neural Network (DNN)}
\begin{figure}[t]
 \centerline{\includegraphics[width=1\linewidth]{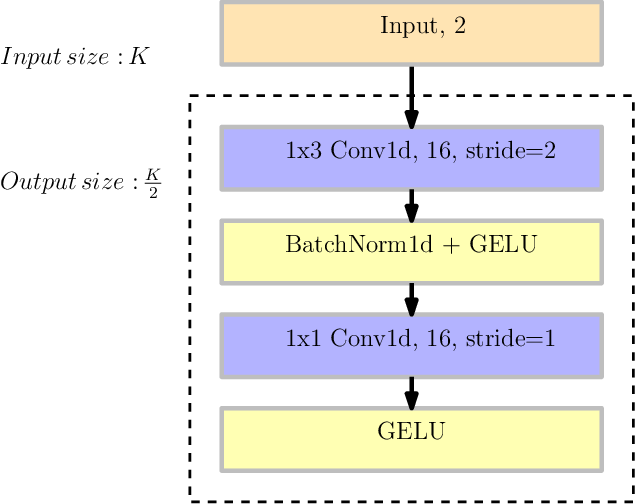}}
  \caption{Basic block structure (dashed part) with 2 input channels and 16 output channels, $\mathrm{BB (2,16)}$.}
  \label{BasicBlock}
  \vspace*{-0.4cm}
\end{figure}
\subsection{Deep Neural Network (DNN) Architecture}
In this section, we present a DNN framework to have end-to-end beamforming design $\{\tilde{w_i} \}_{i=1}^N$ and power allocations $\{p_i\}_{i=1}^N$ for sum-rate maximization problem in (\ref{sum-rate maximization problem}). The DNN input is the frequency domain channel response $\mathbf{H}$ when outputs are beamforming weights $\mathbf{\tilde{W}}$ and power allocations $\mathbf{p}$ as result of joint training procedure. The backbone structure of the proposed DNN architecture is composed of basic blocks (BB) as shown in Fig~\ref{BasicBlock}. BB structure consists of convolutional layers followed by batch normalization and activation layers. The convolutional layers process the frequency domain information obtained by Fourier transform of channel taps. We assume flat fading over time slots, with a maximum Doppler shift of 10 Hz. Thus, changes in channel coefficients are confined to variations across subcarriers. Then, we employ 1D convolutions that operates on the frequency domain. Input data shape is taken as $\left(BNM,2,K\right)$, where $B$ stands for the batch size of MU-MISO channel matrices, and the depth dimension represents the IQ samples, while $K$ represents the number of frequency components. Batch normalization is utilized to provide faster convergence and stability against different initialization of network parameters when GELU activation function is employed since it provides performance improvement compared to RELU and ELU activation functions for different learning fields such as computer vision, speech processing, and natural language processing \cite{hendrycksGELU}.

Moreover, we enhance the number of channels while reducing the size of the feature map within the BB structure. By taking into account the local correlations of physical channels in frequency domain, expanding the depth of the network yields improved representation of latent space. It allows for a more concentrated analysis of local channel characteristics. This strategy is commonly employed in computer vision tasks using popular model architectures to increase the non-linearity, thus enabling the capture of complex relationships within the data \cite{he2016deep,simonyan2015very}.

\begin{figure}[t]
 \centerline{\includegraphics[width=1\linewidth]{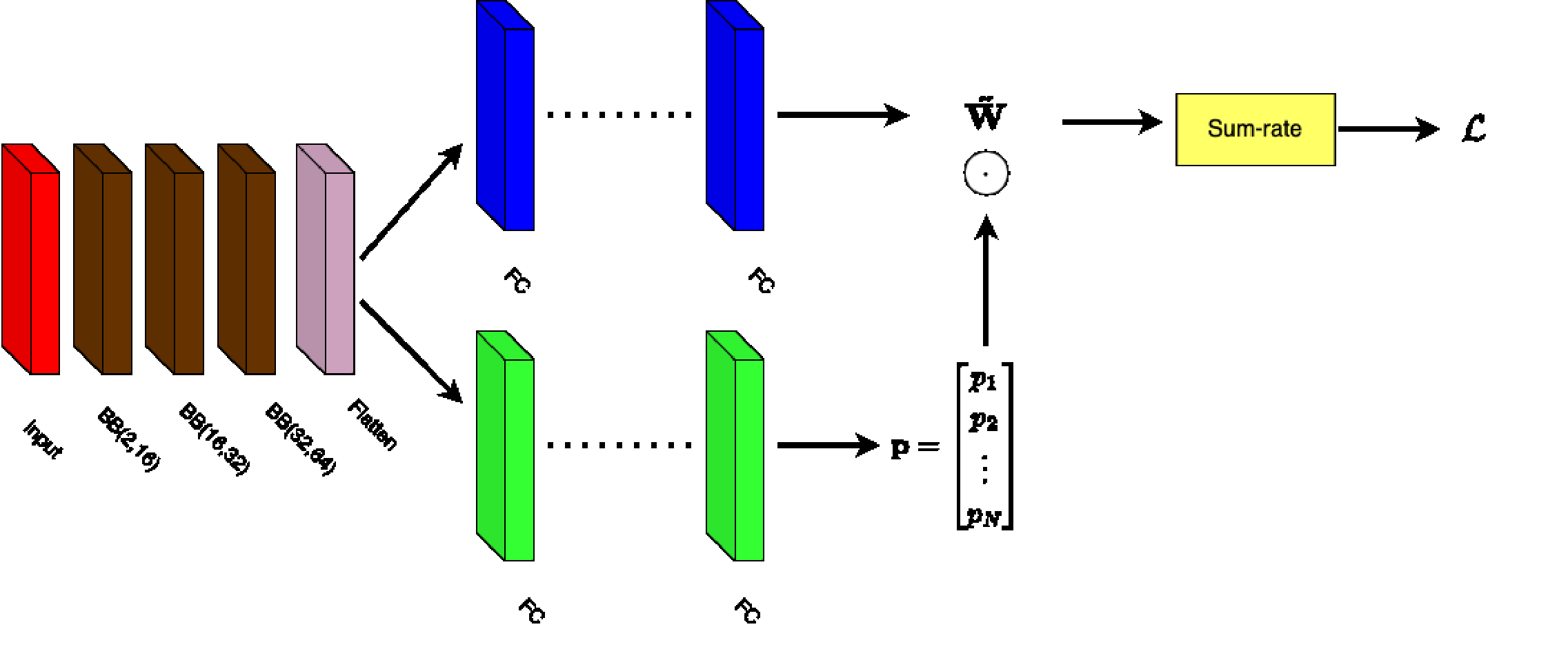}}
  \caption{Deep neural network architecture with joint training procedure.}
  \label{DNN_architecture}
  \vspace*{-0.4cm}
\end{figure}

The DNN architecture is depicted in Fig.~\ref{DNN_architecture}. It is simply the backbone network, which is the concatenation of BB structures, followed by fully connected (FC) layers for beamforming $\mathbf{\tilde{W}}$ design and power allocations $\mathbf{p}$  separately. Blocks in backbone network are characterized by prespecified input and output channel quantities. Flatten layer changes the output shape by concatenating depth dimension for all antenna pairs $(n,m)$, where  $n=1,\ldots,N$ and $m=1,\ldots,M $. Then, the input shape of the first FC layer is $(B,8NMK)$ as shown in Fig.~\ref{DNN_architecture}. The output of FC layers for beamforming design task is reshaped to have beamforming weights $\mathbf{\tilde{W}}$. Softmax activation is employed at the output layer for power allocations $\mathbf{p}$  to satisfy the power constraint given in (\ref{sum-rate maximization problem}). 

\subsection{Training Procedure}
The proposed learning procedure offers unsupervised training based on end-to-end KPIs. The aim is to maximize the sum-rate across all UEs. Therefore, the loss function is specified according to the sum-rate maximization problem given in (\ref{sum-rate maximization problem}),
\begin{align} \label{loss_function}
    \mathcal{L}(\bm{\theta};\mathbf{H}) = -\sum_{i=1}^N \alpha_i \log(1 + \gamma_i)
\end{align}
where $\bm{\theta}$ denotes the set of network parameters for backbone network $\bm{\theta_\textrm{b}}$, FC network parameters of power allocation $\bm{\theta_\textrm{p}}$, and FC network parameters of beamforming design $\bm{\theta_\textrm{w}}$. Note that SINR values $\{ \gamma_i\}_{i=1}^N$ are computed by network input $\mathbf{H}$ and network outputs $f(\bm{\theta}; \mathbf{H}) = \{\Tilde{\mathbf{W}}, \mathbf{p} \}$ without any ground truth labels when $f(\cdot)$ denotes the network function. For the performance evaluation of the proposed network, ZFBF and MMSE beamforming are considered as baseline techniques, which can be computed by the channel $\mathbf{H}$ and the noise variance $\sigma^2$ as,
\begin{align}\label{zfbf_formula}
    \mathbf{W}_{zf} &= \left(\mathbf{H}^H\mathbf{H}\right)^{-1}\mathbf{H}^H \\
    \mathbf{W}_{mmse} &= \left(\mathbf{H}^H\mathbf{H}+\sigma^2 \mathbf{I}_N \right)^{-1} \mathbf{H}^H 
\end{align}

\section{Experiments}
In our experiments, we asses the performance of the proposed framework compared to ZFBF and MMSE. As experiment settings, we consider different antenna configurations, modulation types, channel delay profiles and delay spread values. We evaluate the experiment results across SNR range of [-15, 50] dB. Channel responses for dataset generation are created according to the channel delay profile specifications by 3GPP TR 38.901\cite{3gppTR38901}. We create diverse channel conditions across UEs by defining different channel SNR values. SNR jitter is specified as 20 dB when SNR jitter distribution is Gaussian distribution. Other system parameters can be seen in Table~\ref{table:system parameters}. Spectral efficiency is considered as the performance metric. We refer to the work \cite{vahapoglu2023deep} for the advantage of the proposed work in terms of the computational time complexity. Rather, we examine the proposed framework extensively. A comprehensive summary of experiments is exhibited in Table~\ref{table:experiment summary}. 

\begin{table} [t!]
\begin{center}
\resizebox{\columnwidth}{!}{%
\begin{tabular}{| c | c |}
\hline
 \textbf{Parameter} & \textbf{Value} \\ \hline
 Channel delay profile & TDL-A, TDL-C  \\  \hline
Number of resource blocks (RBs) & 4 (48 subcarriers)  \\ \hline
Delay spread (ns) & 30, 100, 300  \\  \hline
Maximum Doppler shift & 10 Hz  \\ \hline
Subcarrier spacing & 30 kHz \\ \hline
Transmission time interval (TTI) & 500 $\mu s$ \\ \hline
SNR & [-15,50] dB   \\\hline
Modulation scheme & QPSK, 16QAM \\ \hline
SNR jitter & 20 dB \\ \hline
SNR jitter distribution & Gaussian \\ \hline
\end{tabular}}
\end{center}
\caption{System parameters.}
\label{table:system parameters}
\vspace*{-0.5cm}
\end{table}

\subsection{Results and Analysis}
\begin{figure}[t]
 \centerline{\includegraphics[width=.85\linewidth]{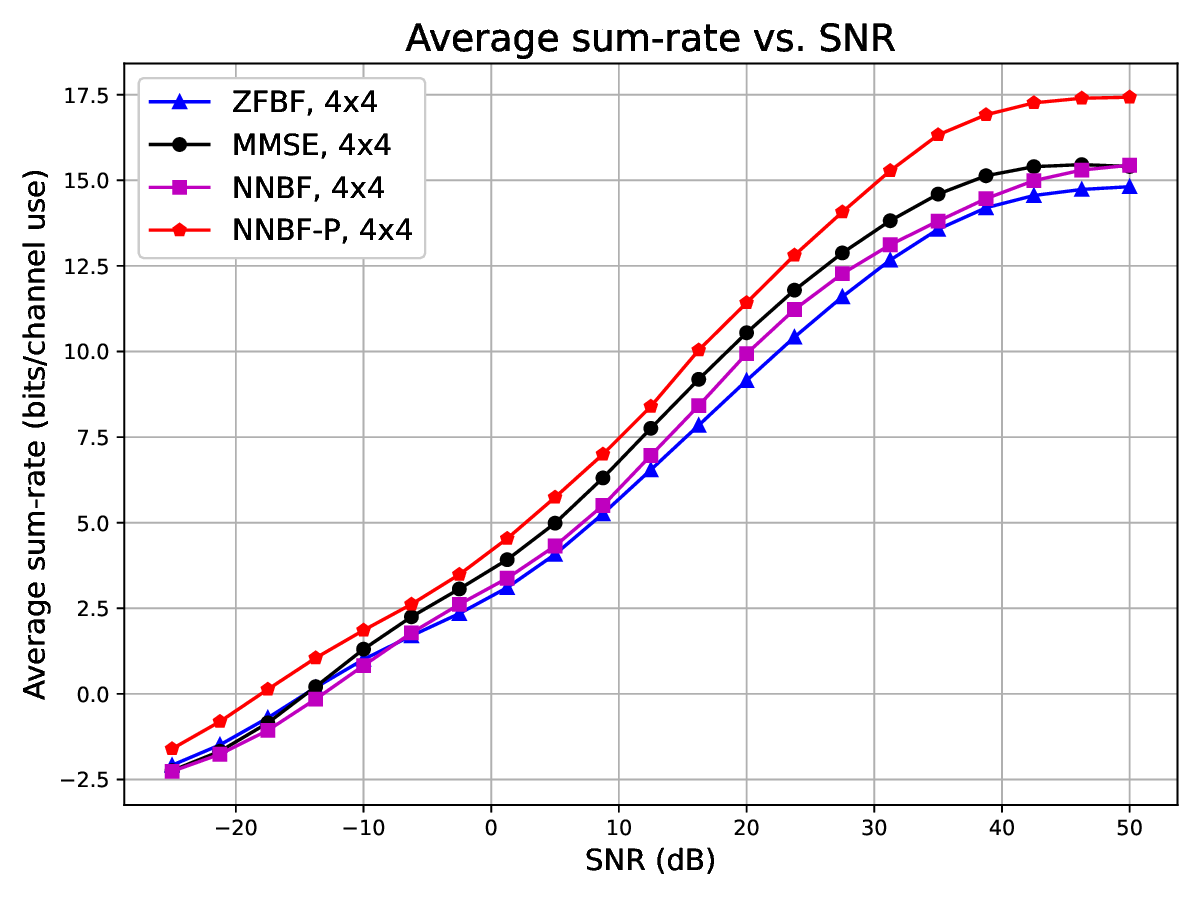}}
  \caption{Performance comparison of NNBF and NNBF-P with baseline methods ZFBF and MMSE when the channel delay profile is TDL-C with delay spread of $300$ ns and modulation type is 16QAM for $M=4$, $N=4$.}
  \label{fig:4x4 baseline comparison}
\end{figure}

In this section, NNBF-P denotes the joint power allocation and beamforming design when NNBF performs beamforming design without power allocation. Therefore, it is considered that $p_i$ is $\frac{P_\textrm{max}}{N}, \forall i$ for ZFBF, MMSE, and NNBF and  $\lambda_i$ is $\frac{P_\textrm{max}}{N}, \forall i$ for MMSE.
Fig.~\ref{fig:4x4 baseline comparison} illustrates the performance comparison for ZFBF, MMSE, NNBF, and NNBF-P when the channel delay profile is TDL-C with the delay spread of 300 ns and the modulation type is 16QAM for $M=4$, $N=4$. MMSE performs better ZFBF and NNBF when ZFBF and NNBF are comparable. The proposed framework NNBF-P considerably surpasses the performances of ZFBF, MMSE, and NNBF for all range of SNR. It shows the significance of joint power allocation scheme. Fig.~\ref{fig:4x4 baseline comparison} corresponds to experiment $10$ in Table~\ref{table:experiment summary}. SINR gain achieved by NNBF-P can also be seen in Table~\ref{table:experiment summary} for specific SNR values. 

\FloatBarrier

\begin{figure}[t]
    \begin{center}
     	\subfigure[]{%
     	\includegraphics[width=0.48\linewidth]{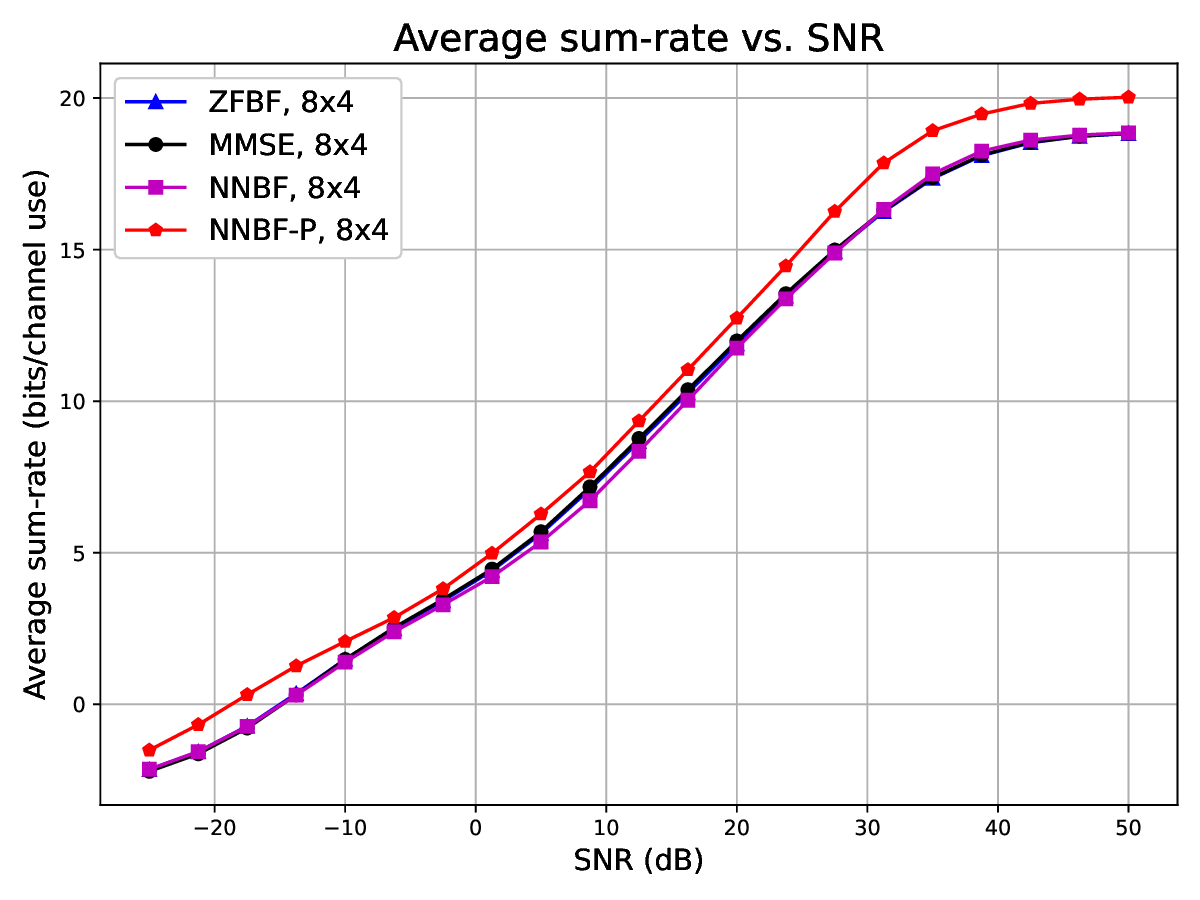}}
        \subfigure[]{%
     	\includegraphics[width=0.48\linewidth]{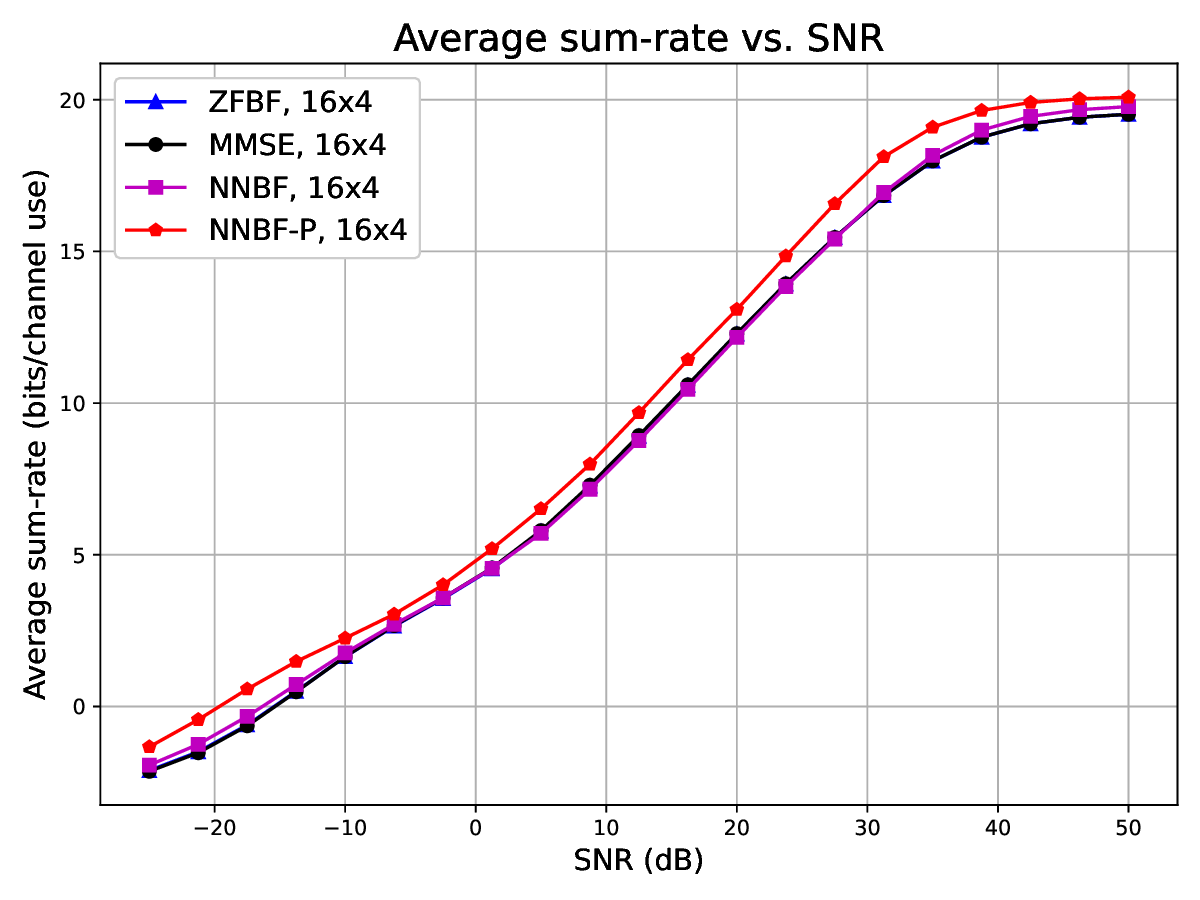}}

    \end{center}
     \caption{Performance comparison of NNBF and NNBF-P with baseline methods ZFBF and MMSE when the channel delay profile is TDL-C with delay spread of $300$ ns and modulation type is 16QAM for different antenna configurations: (a) $M=8$, $N=4$ (b) $M=16$, $N=4$.}
    \label{fig:8x4 and 16x4 baseline comparison}
\end{figure}

Similarly, Fig.~\ref{fig:8x4 and 16x4 baseline comparison} shows the superiority of the proposed framework compared to ZFBF, MMSE, and NNBF for $M=\{8,16\}$ and $N=4$ when NNBF is comparable with ZFBF and MMSE as $M$ increases. They can be seen as experiment $11$ and experiment $12$ in Table~\ref{table:experiment summary}.


\begin{figure}[t]
 \centerline{\includegraphics[width=.85\linewidth]{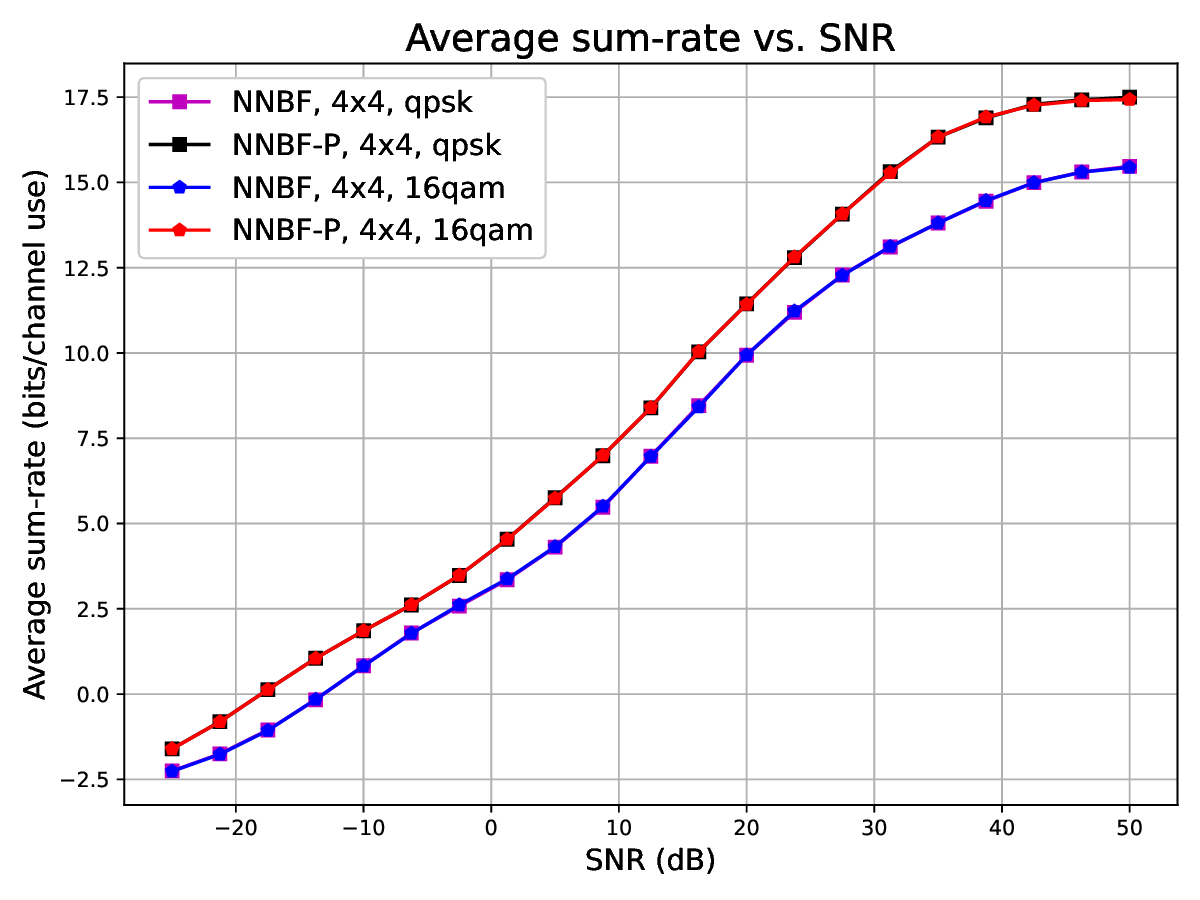}}
  \caption{Performance comparison of NNBF and NNBF-P for QPSK and 16QAM when TDL-C is the channel delay profile with delay spread of $300$ ns and $M=4$, $N=4$.}
  \label{fig: modulation comparison}
\end{figure}

Fig.~\ref{fig: modulation comparison} compares the performances of NNBF and NNBF-P for QPSK and 16QAM. It can be seen that the proposed framework provides similar performance as the order of modulation increases. It shows the robustness of the proposed framework for higher order modulations.

\FloatBarrier

\begin{figure}[t]
    \begin{center}
     	\subfigure[]{%
     	\includegraphics[width=0.48\linewidth]{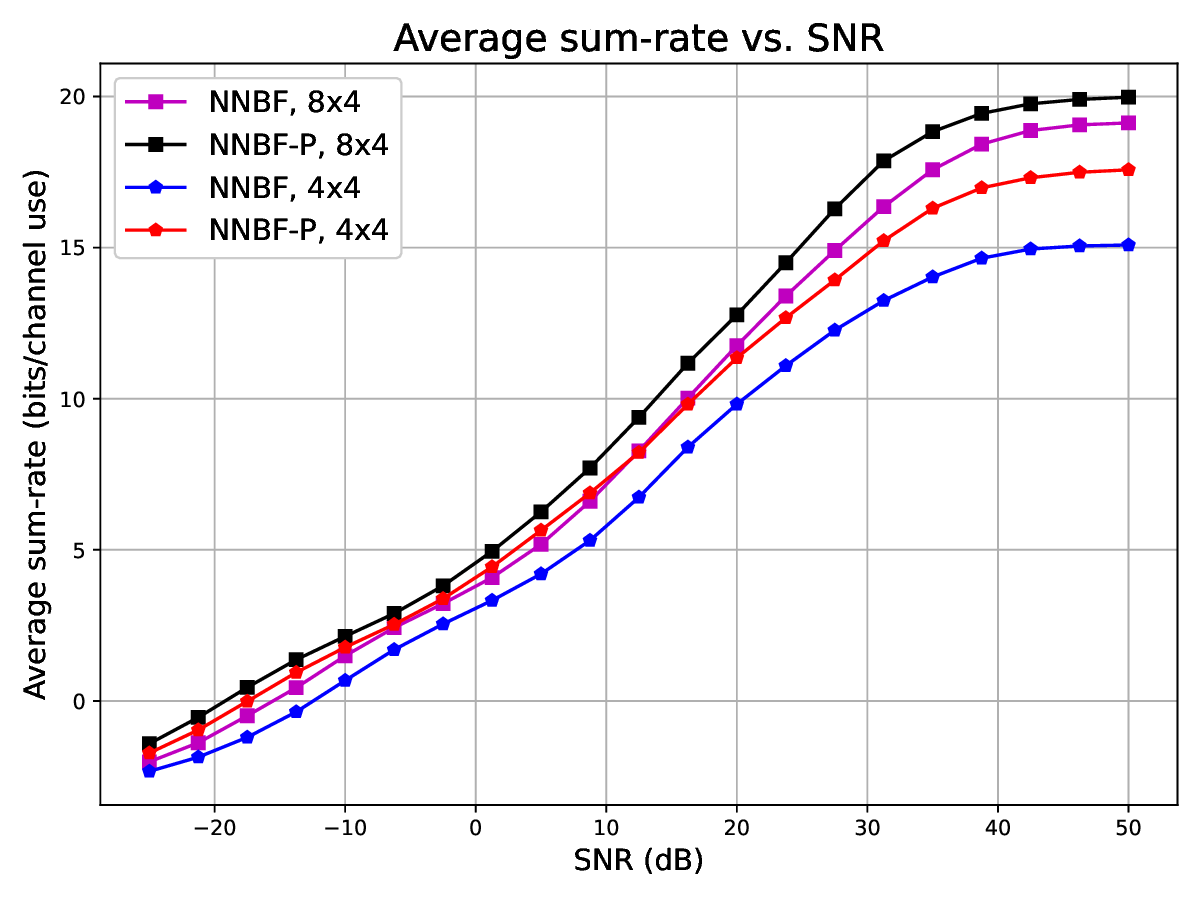}}
     	\subfigure[]{%
     	\includegraphics[width=0.48\linewidth]{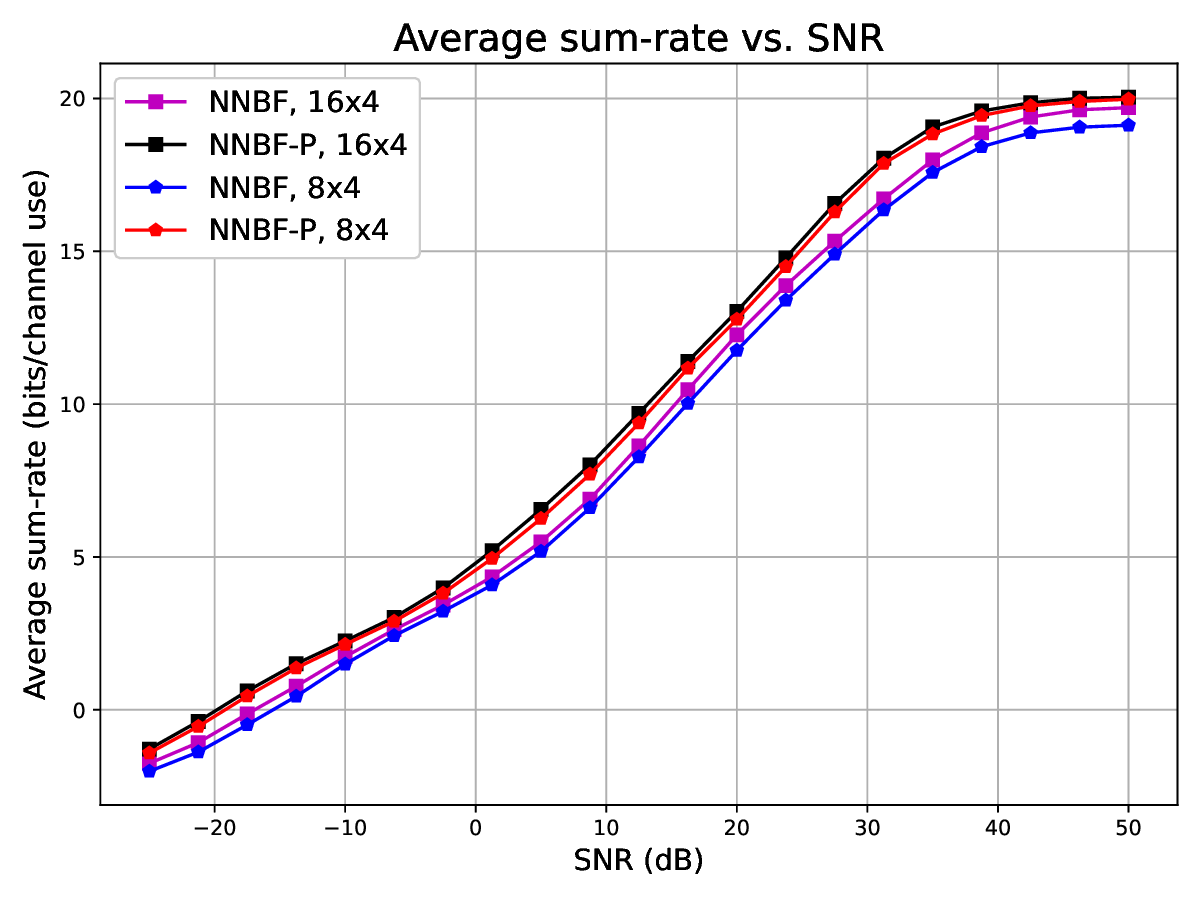}}

    \end{center}
     \caption{Performance comparison of NNBF and NNBF-P when the channel delay profile is TDL-A with delay spread of $30$ ns and modulation type is qpsk for different antenna configurations: (a) $8\times4$ vs. $4\times4$ (b) $16\times4$ vs. $8\times4$.}
    \label{fig: allocated vs. equal}
\end{figure}

Fig.~\ref{fig: allocated vs. equal} compares NNBF and NNBF-P when the channel delay profile is TDL-A with delay spread of 30 ns and modulation type is QPSK. The first subfigure exhibits the comparison of NNBF and NNBF-P for $8\times4$ and $4\times4$ antenna configurations when the second subfigure evaluates NNBF and NNBF-P results for $16\times4$ and $8\times4$ antenna configurations across channel SNR values. It can be seen that NNBF-P for $4\times4$ (red pentagon in subfigure a) is competitive with NNBF for $8\times4$ (magenta square in subfigure a) on low and moderate SNR regimes. Additionally, NNBF-P for $8\times4$ (red pentagon in subfigure b) provides higher results than NNBF for $16\times4$ on all SNR regimes. These results show the success of power allocation scheme achieving competitive or better performance with less antenna equipment.

\begin{table*} [t]
\begin{center}

\resizebox{\linewidth}{!}{%
\renewcommand{\arraystretch}{0.25}
\begin{tabular}{|c | c c c c| c|} \toprule

 \textbf{Experiment no} & \textbf{Channel delay profile} & Delay spread (ns) & Modulation & $M\times N$ & SINR (bps)\\ 
 \midrule\midrule[.1em]
 
     \multirow{2}{*}{1} & \multirow{2}{*}{TDL-A} & \multirow{2}{*}{30} & \multirow{2}{*}{QPSK} & \multirow{2}{*}{$4\times4$} &   
     2.35 $|$ 3.04 $|$ 2.55 $|$ 3.38 \\ \cline{6-6}
     & & & & & 4.16 $|$ 4.96 $|$ 4.20 $|$ 5.64 \\ \midrule

    \multirow{2}{*}{2} & \multirow{2}{*}{TDL-A} & \multirow{2}{*}{30} & \multirow{2}{*}{QPSK} & \multirow{2}{*}{$8\times4$} &   
    3.44 $|$ 3.46 $|$ 3.21 $|$ 3.81 \\ \cline{6-6}
    & & & & & 5.10 $|$ 5.53 $|$ 5.18 $|$ 6.25 \\ \midrule
    
    \multirow{2}{*}{3} & \multirow{2}{*}{TDL-A} & \multirow{2}{*}{30} & \multirow{2}{*}{QPSK} & \multirow{2}{*}{$16\times4$} &   
    3.50 $|$ 3.51 $|$ 3.42 $|$ 3.98 \\ \cline{6-6}
    & & & & & 5.57 $|$ 5.58 $|$ 5.49 $|$ 6.55 \\ \midrule


      \multirow{2}{*}{4} & \multirow{2}{*}{TDL-C} & \multirow{2}{*}{100} & \multirow{2}{*}{QPSK} & \multirow{2}{*}{$4\times4$} &   
      2.33 $|$ 3.10 $|$ 2.66 $|$ 3.53 \\ \cline{6-6}
      & & & & & 4.05 $|$ 5.00 $|$ 4.27 $|$ 5.79 \\ \midrule

     \multirow{2}{*}{5} & \multirow{2}{*}{TDL-C} & \multirow{2}{*}{100} & \multirow{2}{*}{QPSK} & \multirow{2}{*}{$8\times4$} &   
      3.44 $|$ 3.47 $|$ 3.22 $|$ 3.81 \\ \cline{6-6}
      & & & & & 5.64 $|$ 5.70 $|$ 5.29 $|$ 6.26 \\ \midrule

      \multirow{2}{*}{6} & \multirow{2}{*}{TDL-C} & \multirow{2}{*}{100} & \multirow{2}{*}{QPSK} & \multirow{2}{*}{$16\times4$} &   
      3.56 $|$ 3.59 $|$ 3.57 $|$ 4.00 \\ \cline{6-6}
      & & & & & 5.76 $|$ 5.79 $|$ 5.72 $|$ 6.50 \\ \midrule


      \multirow{2}{*}{7} & \multirow{2}{*}{TDL-C} & \multirow{2}{*}{300} & \multirow{2}{*}{QPSK} & \multirow{2}{*}{$4\times4$} &   
      2.35 $|$ 3.07 $|$ 2.58 $|$ 3.48 \\ \cline{6-6}
      & & & & & 4.08 $|$ 4.98 $|$ 4.31 $|$ 5.76 \\ \midrule

     \multirow{2}{*}{8} & \multirow{2}{*}{TDL-C} & \multirow{2}{*}{300} & \multirow{2}{*}{QPSK} & \multirow{2}{*}{$8\times4$} &   
      3.42 $|$ 3.47 $|$ 3.23 $|$ 3.81 \\ \cline{6-6}
      & & & & & 5.64 $|$ 5.70 $|$ 5.31 $|$ 6.27 \\ \midrule

      \multirow{2}{*}{9} & \multirow{2}{*}{TDL-C} & \multirow{2}{*}{300} & \multirow{2}{*}{QPSK} & \multirow{2}{*}{$16\times4$} &   
      3.56 $|$ 3.59 $|$ 3.57 $|$ 4.00 \\ \cline{6-6}
      & & & & & 5.77 $|$ 5.80 $|$ 5.73 $|$ 6.51 \\ \midrule


      \multirow{2}{*}{10} & \multirow{2}{*}{TDL-C} & \multirow{2}{*}{300} & \multirow{2}{*}{16QAM} & \multirow{2}{*}{$4\times4$} &   
      2.34 $|$ 3.06 $|$ 2.61 $|$ 3.48 \\ \cline{6-6}
      & & & & & 4.08 $|$ 4.98 $|$ 4.32 $|$ 5.74 \\ \midrule

     \multirow{2}{*}{11} & \multirow{2}{*}{TDL-C} & \multirow{2}{*}{300} & \multirow{2}{*}{16QAM} & \multirow{2}{*}{$8\times4$} &   
      3.41 $|$ 3.46 $|$ 3.28 $|$ 3.81 \\ \cline{6-6}
      & & & & & 5.63 $|$ 5.69 $|$ 5.35 $|$ 6.27 \\ \midrule

      \multirow{2}{*}{12} & \multirow{2}{*}{TDL-C} & \multirow{2}{*}{300} & \multirow{2}{*}{16QAM} & \multirow{2}{*}{$16\times4$} &   
      3.56 $|$ 3.58 $|$ 3.58 $|$ 4.00 \\ \cline{6-6}
      & & & & & 5.77 $|$ 5.80 $|$ 5.70 $|$ 6.51 \\ \midrule

\end{tabular}}

\end{center}
\caption{Experiment Results Summary for SNR = -2.5 dB and SNR = 5 dB}
\label{table:experiment summary}
\vspace*{-0.5cm}
\end{table*}

\bibliographystyle{unsrt}
\bibliography{main}
\end{document}